\documentclass[prc,twocolumn,showpacs,showkeys]{revtex4}
\usepackage{epsf}
\usepackage{amssymb}
\usepackage{dcolumn}
\def\g{\gamma}
\def\m{\mu^2}
\def\G{\Gamma}

\def\i{\prime}
\def\am{(\alpha_1-\beta_1)}
\def\ap{(\alpha_1+\beta_1)}
\def\amn{\am_{\pi^0}}
\def\amc{\am_{\pi^{\pm}}}
\def\apc{\ap_{\pi^{\pm}}}
\def\apn{\ap_{\pi^0}}
\def\bm{(\alpha_2-\beta_2)}
\def\bp{(\alpha_2+\beta_2)}
\def\bmn{\bm_{\pi^0}}

\def\bpc{\bp_{\pi^{\pm}}}
\def\bpn{\bp_{\pi^0}}
\def\gg{\g \g\to \pi^0 \pi^0}
\def\s{\sigma}
\def\sig{\frac{d\s_{\gg}}{d\Omega}}
\def\tg{\theta^{*}}

\def\mpp{M_{++}}
\def\mm{M_{+-}}
\def\sp{s'}
\def\tp{t'}
\def\be{\begin{equation}}
\def\ee{\end{equation}}
\def\beq{\begin{eqnarray}}
\def\eeq{\end{eqnarray}}
\begin{document}

\title{Determination of $\pi^0$ meson quadrupole polarizabilities from
the process\\
$\gg$}
\vskip 0.5cm

\author{L.V.~Fil'kov}
\email[E-mail: ]{filkov@sci.lebedev.ru}
\author{V.L.~Kashevarov}
\email[E-mail: ]{kashev@kph.uni-mainz.de}
\affiliation{P.N. Lebedev Physical Institute, Leninsky Prospect 53,
Moscow 119991, Russia}
\vskip 1cm

\begin{abstract}
A fit of the experimental data to the total cross section of the
process $\gg$ in the energy region from 270 to 2250 MeV
has been carried out using dispersion relations for the
invariant amplitudes where the quadrupole polarizabilities are free
parameters. As a result the sum and difference of the electric and
magnetic quadrupole polarizabilities of the $\pi^0$ meson have been found
for the first time: \mbox{$\bpn=(-0.181\pm 0.004)\times 10^{-4}$\,fm$^5$},
\mbox{$\bmn=(39.70\pm 0.02)\times 10^{-4}$\,fm$^5$}. In addition,
dispersion sum rules have been constructed for this sum and difference,
respectively.
The values of $\bmn$ and $\bpn$ extracted from the experimental data on
the process $\gg$ are in good agreement with the result of calculations
in the framework of these dispersion sum rules.
\end{abstract}
\vskip 0.3cm


\maketitle

\section{Introduction}

Hadron polarizabilities are structure parameters, the values of which are very
sensitive to predictions of different theoretical models. Therefore, accurate
experimental determination of these values
provides a method for testing the validity of
such models. However, because there are
no pion targets, it is necessary
to use indirect methods to determine the pion polarizabilities.
For example, the dipole polarizabilities of charged pions can be determined
either from the scattering of the high energy pions off the Coulomb field of
heavy nuclei \cite{ant,pom,star} or from radiative $\pi^+$ photoproduction
from the proton \cite{ayb,drech,ahr}. Unfortunately, the results obtained
from analyses of the reaction $\g\g\to\pi\pi$ at low energies
\cite{bab,kal,don} are essentially model dependent due to the strong $S$-wave
$\pi \pi$ interaction in this energy region.

Moreover, due to the fact that
the Born term for the reaction $\g\pi^0\to\g\pi^0$ is equal to zero,
extraction of the $\pi^0$ meson polarizabilities by
extrapolating the experimental data on the radiative $\pi^0$
photoproduction from the proton to the pion pole is ineffective.
At present the most reliable method in this case is an analysis of the
process $\gg$ in the region of the $f_2(1270)$ meson where
the cross section of this process is very sensitive to the values of the
$\pi^0$ polarizabilities.
%the contribution of the $\pi^0$ meson polarizabilities is big enough.
In the work \cite{ser} the analysis of the angular distributions of pions from
the process $\gg$ in this energy region has resulted in $\ap =1.00\pm 0.05$
(in units of $10^{-4}$\,fm$^3$).

The fit of the data \cite{mars} for the total cross section of the
process $\gg$, using dispersion relations (DR) at fixed $t$
(where $t$ is the square of the total energy in $\g\g$ c.m.s.) with one
subtraction for the invariant amplitudes \cite{fil1} in the energy region
from 270 up to 2000 MeV, has allowed for the determination of $\pi^0$ dipole
polarizabilities $\apn=0.98\pm 0.03$ and $\amn=1.6\pm 2.2$.
Here the $\s$ meson was considered as an effective description of
the strong $S$-wave $\pi\pi$ interaction using the broad Bright-Wigner
resonance expression. The parameters of such a $\s$ meson were found
from the fit to the experimental data \cite{mars} in the energy region
270--825 MeV.
As a result, a good description of the experimental data \cite{mars}
was obtained for $\sqrt{t}=$270--1700 MeV.
However, this model predicts a strong rise in the total cross section
at higher energies in contradiction with the experimental data.

In the present work we show that this discrepancy can be eliminated in the
energy region at least up to 2.25 GeV by considering the quadrupole $\pi^0$
meson polarizabilities as free parameters.

An investigation of the process $\gg$ at low and middle energies was also
carried out
%in a number of works
in the framework of different theoretical
models \cite{bij,bell,oller,lee}. However, these models did not allow a
good description of the experimental data \cite{mars} on the total
cross section in the full energy region from 270 to 2000 MeV. In the present
work good agreement with the experimental data under
consideration has been obtained in the full energy region from 270 to
2250 MeV.

The quadrupole polarizabilities of pions and nucleons were investigated
in Ref.~\cite{rad1} where, in particular, the sum of the electric and
magnetic polarizabilities of the pions and the sum and difference 
for the proton have been estimated for the first time using
dispersion sum rules. In Ref.~\cite{lvov,holst} the quadrupole
polarizabilities of the nucleons were calculated with help of dispersion
relations and the results obtained were compared to predictions based upon
chiral symmetry.

As was shown in Ref.~\cite{fil2,drech}, the quadrupole
polarizabilities give a big contribution to
the cross section of Compton scattering on the $\pi^0$.

In the present paper, the contribution of the quadrupole polarizabilities
of the $\pi^0$ meson to the total cross section of the process $\gg$ is
studied  using dispersion relations at fixed $t$
with a subtraction for invariant amplitudes. The subtraction functions
are determined by the DRs in cross channels with two subtractions
where the subtraction constants are connected with dipole and quadrupole pion
polarizabilities. The subtractions in the DRs provide good convergence
of the integrand expressions of these DRs and so increase the reliability
of the calculations.

It is shown that the total cross section of the process $\gg$ is very
sensitive to the values of the quadrupole polarizabilities in the energy
region higher than 1250 MeV. The fit of the experimental data \cite{mars,bien}
to the process $\gg$ using the DRs constructed has
allowed for the determination of the values of the sum and the difference of
the $\pi^0$ meson quadrupole polarizabilities for the first time.

In order to analyze these values of the quadrupole polarizabilities,
the dispersion sum rules (DSR) for $\bm$ and $\bp$ are constructed.
The prediction of DSRs and ChPT for the dipole polarizabilities of the
charged and neutral pions are compared with the existing experimental
values.

The paper is organized as follows. In Sec. II DRs for the invariant
amplitudes of the process $\gg$ are constructed. In Sec. III the dispersion
sum rules for the dipole and quadrupole pion polarizabilities are
constructed and analyzed.
The determination of the $\pi^0$ quadrupole polarizabilities from
the experimental data on the process $\gg$ follows in Sec. IV.
The main conclusions are presented in Sec. V. The details of the calculations
of meson resonances are given in the Appendix.

\section{Dispersion relations for the amplitudes of the process $\gg$}

The dipole and quadrupole polarizabilities arise as ${\cal O}(\omega^2)$
and ${\cal O}(\omega^4)$ terms, respectively, in the expansion of the
non-Born amplitude of Compton scattering over the initial photon energy
$\omega$. In terms of the electric $\alpha_l$ ($l=$1, 2) and magnetic
$\beta_l$ dipole and quadrupole polarizabilities, the corresponding effective
interactions of ${\cal O}(\omega^2)$ and
${\cal O}(\omega^4)$ have the forms \cite{lvov}:
\be
H^{(2)}_{eff}=-\frac12\,4\pi\,(\alpha_1\,\vec{E}^2+\beta_1\,\vec{H}^2),
\ee
\be
H^{(4)}_{eff}=-\frac{1}{12}4\pi\left(\alpha_2\,E^2_{ij}+
\beta_2\,H^2_{ij}\right)
\ee
where
\be
E_{ij}=\frac12\,\left(\nabla_i\,E_j+\nabla_j\,E_i\right), \;
H_{ij}=\frac12\,\left(\nabla_i\,H_j+\nabla_j\,H_i\right)
\ee
are the quadrupole strengths of the electric and magnetic fields.

The dipole polarizabilities ($\alpha_1$ and $\beta_1$) of the pion
measure the response of the pion to quasistatic electric and magnetic
fields. On the other hand, the parameters $\alpha_2$ and $\beta_2$
measure the electric and magnetic quadrupole moments induced in the pion
in the presence of an applied field gradient.

In order to determine the quadrupole polarizabilities of $\pi^0$ meson,
we will consider the process $\gg$.
This process is described by the following invariant
variables
\be
t=(k_1+k_2), \quad s=(p_1-k_1)^2, \quad u=(p_1-k_2)^2
\ee
where $p_1(p_2)$ and $k_1(k_2)$ are the pion and photon
4-momenta.

We will consider the helicity
amplitudes $\mpp$ and $\mm$ \cite{aber} which are expressed through Prange's
amplitudes \cite{pran} $T_1$ and $T_2$ as
\beq
\mpp&=&-\frac{1}{2t}\left(T_1+T_2\right), \nonumber \\
\mm&=&-\frac{T_1-T_2}{2[(s-\m)^2+st]}.
\label{mm}
\eeq
These amplitudes have no kinematical singularities or zeros
and define the cross section of the process $\gg$
as follows
\beq
\sig&=&\frac1{256\pi^2}\sqrt{\frac{(t-4\m)}{t^3}}\left\{t^2 |\mpp|^2
\right. \nonumber \\
&& \left. +\frac{1}{16}t^2(t-4\m)^2\sin^4\tg|\mm|^2\right\}
\label{sig}
\eeq
where $\tg$ is the angle between the photon and the pion in the c.m.s.
of the process $\gg$ and $\mu$ is the $\pi^0$ meson mass.

Constructing the DR at fixed $t$ with one subtraction at $s=\m$
for the amplitude $\mpp$ we have:
\beq
&&Re \mpp (s,t)=Re \mpp (s=\m,t) \nonumber \\
&&+\frac{(s-\m)}{\pi}P~\int\limits_{4\m}^{\infty}d\sp~Im\mpp(\sp,t)\left[
\frac{1}{(\sp-s)(\sp-\m)}\right. \nonumber \\
&&\left.-\frac{1}{(\sp-u)(\sp-\m+t)}\right].
\label{dr1}
\eeq
Via the crossing symmetry this DR is identical to a DR with two subtractions.

We determine the subtraction function $Re \mpp (s=\m,t)$
with the help of the DR at fixed $s=\m$ with two subtractions using
the crossing symmetry between the $s$ and $u$ channels
\beq
%\lefteqn{
&&Re \mpp(s=\m,t)= \nonumber \\
&&=\mpp (s=\m,0)+\left.t\frac{d\mpp (s=\m,t)}{dt}
\right|_{t=0} \nonumber \\
&& +\frac{t^2}{\pi}\left\{P\int\limits_{4\m}^{\infty}
\frac{Im\mpp(\tp,s=\m)~d\tp}{\tp^2(\tp-t)}\right. \nonumber \\
&&\left. +\int\limits_{4\m}^{\infty}
\frac{Im\mpp(\sp,u=\m)~d\sp}{(\sp-\m)^2(\sp-\m+t)}\right\}.
\label{sub}
\eeq

Taking into account the expressions of the sum and the difference of
the generalized electric and magnetic polarizabilities of any multipolar
order through invariant amplitudes \cite{guias}, we determine the
subtraction constants $\mpp(s=\m,t=0)$ and
$\left.d\mpp (s=\m,t)/dt\right|_{t=0}$ in terms of differences of
the dipole ($\amn$) and quadrupole ($\bmn$) polarizabilities

\beq
&&\mpp(s=\m,t=0)=2\pi\mu\amn, \nonumber \\
&&\left.\frac{d\mpp(s=\m,t)}{dt}\right|_{t=0}=\frac{\pi\mu}{6}\bmn.
\label{a-b}
\eeq
%where $\alpha_l$ and $\beta_l$ are the electric and magnetic pion
%polarizabilities, respectively.

The DRs for the amplitude $\mm(s,t)$ have the same expressions
(\ref {dr1}) and (\ref{sub}) but
with the substitutions: $\mpp\to \mm$ and $Im\mpp \to Im\mm$.
The subtraction constants are equal in this case to
\beq
&&\mm(s=\m,t=0)=\frac{2\pi}{\mu}\apn, \nonumber \\
&&\left.\frac{d\mm(s=\m,t)}{dt}\right|_{t=0}=\frac{\pi}{6\mu}\bpn .
\label{a+b}
\eeq

\section{Dispersion sum rules for the pion polarizabilities}

The DSR for the difference of the dipole polarizabilities was obtained
in Ref. \cite{rad} using DR at fixed $u=\m$ without subtractions
for the amplitude $\mpp$:
\beq
\am&=&
\frac{1}{2\pi^2\mu}\left\{\int\limits_{4\m}^{\infty}~\frac{
Im\mpp(\tp,u=\m)~d\tp}{\tp}\right. \nonumber \\
&&\left. +\int\limits_{4\m}^{\infty}~\frac{
Im\mpp(\sp,u=\m)~d\sp}{\sp-\m}\right\}.
\label{dsr1m}
\eeq
The DSR for the sum of the dipole polarizabilities reads
\beq
%\lefteqn{
\ap&=&\frac{\mu}{\pi^2}~\int\limits_{4\m}^{\infty}~\frac{
Im\mm(\sp,t=0)~d\sp}{\sp-\m} \nonumber \\
&=&\frac{1}{2\pi^2}~\int\limits_{\frac32\mu}^{\infty}~\frac{
\s_T(\nu)d\nu}{\nu^2}
\label{dsr1p}
\eeq
where $\s_T$ is the total cross section of the $\g\pi$ interaction
and $\nu$ is the photon energy in the lab. system.

The DSRs for the difference and the sum of the
quadrupole polarizabilities
can be obtained with the help of the DRs at fixed $u=\m$ with one
subtraction for the amplitudes $\mpp$  and $\mm$, respectively:
\beq
\bm&=&
\frac{6}{\pi^2\mu}\left\{\int\limits_{4\m}^{\infty}~\frac{
Im\mpp(\tp,u=\m)~d\tp}{\tp^2} \right.\nonumber \\
&&\left. -\int\limits_{4\m}^{\infty}~\frac{
Im\mpp(\sp,u=\m)~d\sp}{(\sp-\m)^2}\right\},
\label{dsrm}
\eeq
\beq
\bp&=&
\frac{6\mu}{\pi^2}\left\{\int\limits_{4\m}^{\infty}~\frac{
Im\mm(\tp,u=\m)~d\tp}{\tp^2}\right.\nonumber \\
&&\left. -\int\limits_{4\m}^{\infty}~\frac{
Im\mm(\sp,u=\m)~d\sp}{(\sp-\m)^2}\right\}.
\label{dsrp}
\eeq

The DSRs for the charged pions are saturated  by the contributions of
the $\rho(770)$, $b_1(1235)$, $a_1(1270)$, and $a_2(1320)$ mesons in
the $s$ channel and $\s$, $f_0(980)$, $f_0^{\i}(1370)$, $f_2(1270)$, and
$f_2^{\i}(1525)$ in the $t$ channel.
For the $\pi^0$ meson the contribution from the $\rho$, $\omega(782)$, and
$\phi(1020)$ mesons are considered in the $s$ channel and from the
same mesons as for the charged pions in the $t$ channel.

The parameters of the $\rho$, $\omega$, $\phi$, $b_1$, $a_2$, $f_2$, and
$f_2^{\i}$ mesons are given by the Particle Data Group \cite{pdg}.
The parameters of the
$f_0(980)$, $f_0^{\i}(1370)$, and $a_1$ mesons are taken  as follows:

\noindent
$f_0(980)$: $m_{f_0}=980$ MeV \cite{pdg}, $\G_{f_0}=70$ MeV (the average
of the PDG \cite{pdg} estimate),
$\G_{f_0\to \g\g}=0.39\times 10^{-3}$ MeV \cite{pdg},
$\G_{f_0\to \pi\pi}=0.84\,\G_{f_0}$ \cite{anis};

\noindent
$f_0^{\i}(1370)$: $m_{f_0^{\i}}=1434$ MeV \cite{ait}, $\G_{f_0^{\i}}=173$ MeV
\cite{ait},
$\G_{f_0^{\i}\to \g\g}=0.54\times 10^{-5}$ MeV \cite{morg},
$\G_{f_0^{\i}\to \pi\pi}=0.26\,\G_{f_0^{\i}}$ \cite{bugg};

\noindent
$a_1(1270)$: $m_{a_1}=1230$ MeV \cite{pdg}, $\G_{a_1}=$425 MeV (the average
value of the PDG estimate \cite{pdg}),
$\G_{a_1\to \g\pi^{\pm}}=0.64$ MeV \cite{zel}.

For the $\s$ meson we use the values of mass and decay widths
found in Ref. \cite{fil1}: $m_{\s}=547$ MeV, $\G_{\s}=1204$ MeV,
$\G_{\s\to \g\g}=0.62$ keV.

The results of the calculations of the DSRs for the dipole (in units
of $10^{-4}$\,fm$^3$) and quadrupole (in units of $10^{-4}$\,fm$^5$)
polarizabilities are presented in Table I for the charged and in Table II
for the neutral pions.
The contributions of the $f_2(1525)$ meson to the DSRs for $\bpc$ and
for $\bpn$ were found to be very small \mbox{$(\sim 0.0004)$} and were not
included in the tables.
The errors indicated are due to
uncertainties in the parameters of the mesons considered.

The influence of the integration limit in the DSRs (\ref{dsr1m})--(\ref{dsrp})
on results of the calculations was investigated. The analysis has shown
that the integrand expressions for the quadrupole polarizabilities
converge very quickly and the integration up to 2 GeV gives practically 100\%.
For the dipole polarizabilities the integrand expressions converge more
slowly, particularly for their difference. While for the sums of the
dipole polarizabilities, an integration limit of 5 GeV gives about
99\%, the integration in the DSRs for the differences up to 10 GeV
introduces uncertainties of $\sim 1\%$ for $\amc$ and $\sim 5\%$ for
$\amn$. In the present work we performed the integrations up to 10 GeV
for the quadrupole polarizabilities and the sum of the dipole
polarizabilities and up to 100 GeV for the differences of the dipole
polarizabilities.

\begin{table*}
\caption{The DSR predictions for the polarizabilities of the charged
pions in units of $10^{-4}$\,fm$^3$ for the dipole polarizabilities and
$10^{-4}$\,fm$^5$ for the quadrupole polarizabilities.}
\centering
\begin{tabular}{ccccccccccc}\hline
 &$\rho$&$b_1$&$a_1$&$a_2$&$f_0$&$f_0^{\i}$&$\sigma$&$f_2$&$\Sigma$
&$\Delta\Sigma$ \\ \hline
$\am$&-1.15&0.93 &2.26 &1.51 &0.58&0.02&9.45&-&13.60&2.15 \\ \hline
$\ap$&0.063&0.021&0.051&0.031& -  & -  & -  &-&0.166&0.024 \\ \hline
$\bm$&0.78 &-0.25&-0.63&-0.41&0.31&0.01&25.94&-&25.75&7.03 \\ \hline
$\bp$&-0.027&-0.003&-0.011&0.013&-&-&-&0.149&0.121&0.064 \\ \hline
\end{tabular}
\end{table*}

\begin{table*}
\caption{The DSR predictions for the polarizabilities of the $\pi^0$
meson.}
\centering
\begin{tabular}{cccccccccc}\hline
 &$\rho$&$\omega$&$\phi$&$f_0$&$f_0^{\i}$&$\sigma$&$f_2$&$\Sigma$
&$\Delta\Sigma$ \\ \hline
$\am$&-1.58 &-12.56&-0.04 &0.60 &0.02 &10.07& - &-3.49 &2.13 \\ \hline
$\ap$&0.080 &0.721 & 0.001 &  -  &  -  &  -  & - &0.802 &0.035 \\ \hline
$\bm$&1.06  & 9.53 & 0.02 &0.32 &0.01 &28.78& - &39.72  &8.01 \\ \hline
$\bp$&-0.035&-0.284& 0    &  -  &  -  &  -  &0.148&-0.171&0.067 \\ \hline
\end{tabular}
\end{table*}

The investigation within the framework of ChPT in a two loop analysis
${\cal O}(p^6)$ \cite{burgi} has yielded the following values for the dipole
polarizabilities of the charged pions:
\beq\label{dipch}
\amc&=&4.4\pm 1.0, \\
\apc&=&0.3\pm 0.1.
\eeq
For the $\pi^0$ meson ChPT has predicted \cite{bell}
\beq\label{dip0}
\amn&=&-1.90\pm 0.20, \\
\apn&=&1.15\pm 0.30.
\eeq

A recent experiment at the Mainz Microtron MAMI \cite{ahr} has resulted in
\be\label{mainz}
\amc=11.6\pm 1.5_{stat}\pm 3.0_{syst}\pm 0.5_{mod}.
\ee
This value of $\amc$ is close to the result of Ref. \cite{ant}.

The dipole polarizabilities of the $\pi^0$ meson were determined by
investigating the process $\gg$ in Ref. \cite{kal,ser,fil1}. These works
have given: $\apn=1.00\pm 0.05$, $\amn=-0.6\pm 1.8$ \cite{kal,ser} and
$\apn=0.98\pm 0.03$, $\amn=-1.6\pm 2.2$ \cite{fil1}.

As seen from Table I, the DSR results in $\amc=13.60\pm 2.15$. This value of
$\amc$ is in agreement within the errors with the experimental data
(\ref{mainz}) but differs significantly from the ChPT prediction (\ref{dipch}).
On the other hand, the DSR calculations for $\pi^0$ meson dipole
polarizabilities (Table II) are not in conflict within the errors with
both experimental data of Ref. \cite{ser,fil1} and the ChPT predictions
(\ref{dip0}).

\section{Determination of the $\pi^0$ meson quadrupole polarizabilities}

To determine the quadrupole polarizabilities of the $\pi^0$ meson,
we fit experimental data on the process $\gg$ in the region
$\sqrt{t}=270$--2250 MeV. As the data of Ref. \cite{mars} have large
errors in the energy region 1600--2000 MeV and so cannot be correctly used
to determine the cross section behavior at higher energies, we in addition
consider the data of Ref. \cite{bien} in the region of 2000--2250 MeV.
We fit these experimental data using the DRs (\ref{dr1})--(\ref{sub})
for the amplitude $\mpp$ and the corresponding DRs for $\mm$ where the
difference and the sum of the quadrupole polarizabilities are free
parameters. The values of the dipole polarizabilities $\amn$ and
$\apn$ and the parameters of the $\s$ meson are taken from Ref.
\cite{fil1}.
In order to improve the description of the
$f_2(1270)$ meson resonance peak, the effective radius $(r_f)$
and the decay width $\G_{f_2\to\g\g}$ of the meson
are considered as free parameters, too.

As a result, the following values have been found:
\be\label{qsum}
\bpn=(-0.181\pm 0.004)\times 10^{-4}\,{\rm fm}^5,
\ee
\be\label{qdif}
\bmn=39.70\pm 0.02\times 10^{-4}\,{\rm fm}^5,
\ee
and
\be\label{rg}
r_f=0.96\pm 0.01\; {\rm fm}, \quad \G_{f_2\to\g\g}=3.05\pm 0.11\; {\rm keV}.
\ee

Note that the value of $r_f$ determined in Ref. \cite{gray} is
equal to $1.05\pm 0.24$ fm. The value of $\G_{f_2\to\g\g}$ practically
coincides with the one found in \cite{mars} and differs from the value
presented by the Particle Data Group \cite{pdg} ($2.61\pm 0.24$ keV).

Fig. 1 shows a fit to the total cross section of the process
$\gg$ in the energy region up to $\sqrt{t}=2.25$ GeV \cite{mars,bien}
using the values of $\bpn$, $\bmn$, $r_f$, and $\G_{f_2\to\g\g}$
found (the solid curve).
\begin{figure}
\epsfxsize=8cm  % 8.5cm
\epsfysize=9.5cm %  10cm
\centerline{
\epsffile{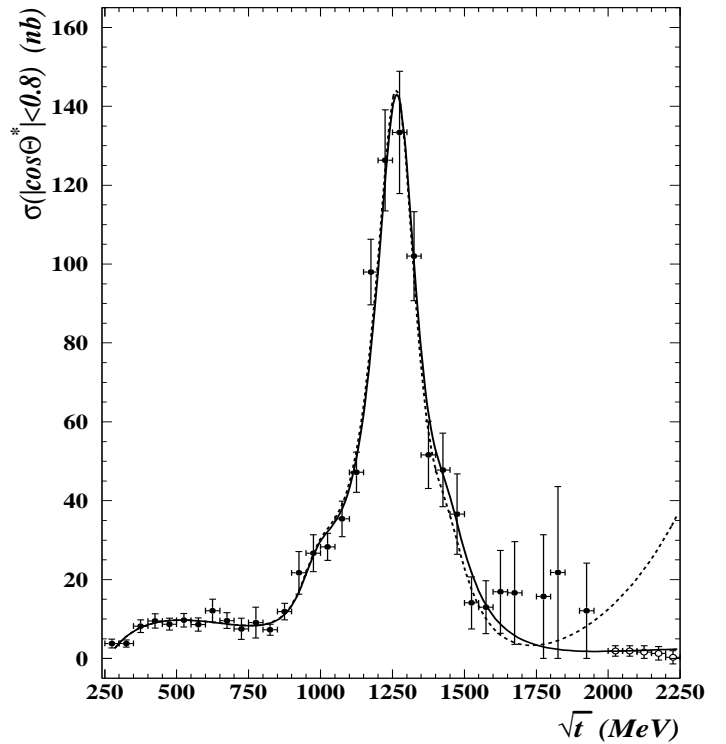}}    %fig1.ps
\caption{The total cross section  of the reaction $\gg$. The solid curve
is the result of the fit. The dashed curve corresponds to
the quadrupole polarizabilities calculated with the help of the DSRs.
The full circles are data from Ref. \cite{mars} and
the open ones are data from Ref. \cite{bien}.}
\end{figure}
The dashed curve corresponds to the values of the quadrupole
polarizabilities obtained with the help of the DSRs (\ref{dsrm}) and
(\ref{dsrp}) using the meson parameters described in Sec. III. As the
predictions of the DSR (\ref{dsrm}) for $\bmn$ effectively coincide with
the values (\ref{qdif}) found in this analysis, the difference between
the predictions and the experiment  in the region of 1750--2250 MeV is
caused by the deviation of $\sim 5\%$ of the DSR
result for $\bpn$ from the experimental value.

The sensitivity of the calculations of the cross section of the process $\gg$
to values of the difference and the sum of the quadrupole polarizabilities
is shown in Fig. 2.
\begin{figure}
\epsfxsize=8.5cm
\epsfysize=10cm
\centerline{
\epsffile{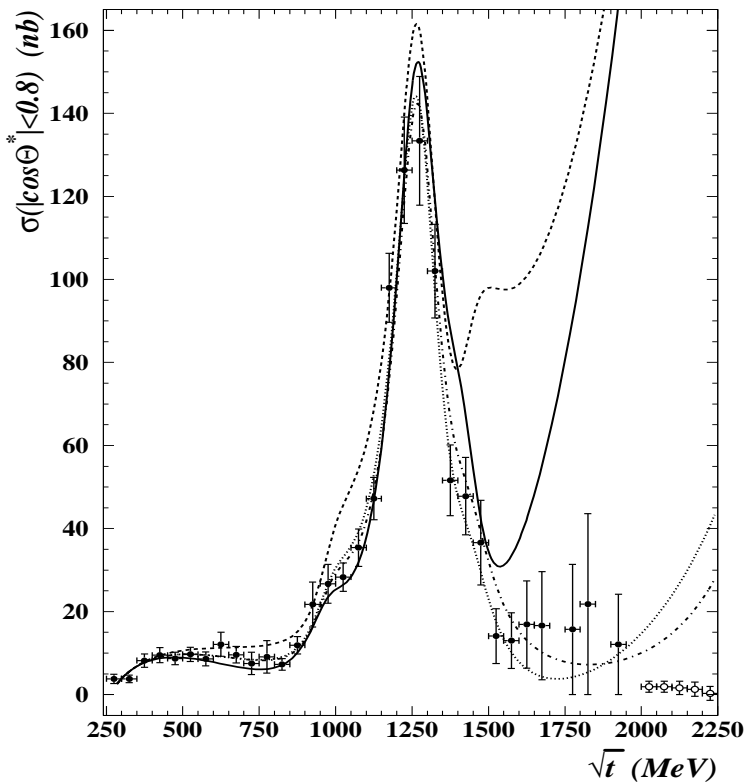}}
\caption{The sensitivity of the cross section calculations to different values
of the quadrupole polarizabilities. The solid (dashed) curve corresponds to
$\bmn$ bigger (less) by 5\% than the experimental value. The dotted
(dashed-dotted) curve presents the result with the same 5\% deviation for
$\bpn$.}
\end{figure}
The solid and dashed curves in this figure present results of the calculations
when the value of $\bmn$ is bigger and less by 5\% than the experimental
value (\ref{qdif}). The dotted and dashed-dotted curves correspond to the
same 5\% deviations from the value (\ref{qsum}) for $\bpn$.

As seen from this figure, the result of the calculations of the cross
section is very sensitive to the value of $\bmn$ in the overall energy
region under
consideration particularly at $\sqrt{t}>1400$ MeV. On the other hand, 5\%
changes of $\bpn$ do not influence the result of the
calculations in the energy region up to 1400 MeV but lead to an
essential difference from the experimental data at the higher energies.

All of this indicates a high sensitivity of our calculations
to the values of the quadrupole $\pi^0$ meson polarizabilities,
particularly in the energy region above 1400 MeV.

The values of $\bmn$ and $\bpn$ we found are consistent within the errors
with the predictions of the DSRs (\ref{dsrm}) and (\ref{dsrp}).

\section{Conclusions}

The DRs at fixed $t$ with one subtraction at $s=\m$ have been constructed
for the invariant amplitudes of the process $\gg$. The subtraction functions
were determined with the help of the DRs at fixed $s=\m$ with two
subtractions at $t=0$, where the subtraction constants were
expressed through the dipole and
quadrupole polarizabilities. These DRs, where the sum and the difference
of the quadrupole $\pi^0$ polarizabilities were free parameters,
were used to fit the
experimental data \cite{mars,bien} for the total cross sections of the
process $\gg$ in the energy region from 270 to 2250 MeV. As a result the
values of the sum and the difference of the quadrupole polarizabilities
have been found for the first time:
$$
\bpn=(-0.181\pm 0.004)\times 10^{-4}\,{\rm fm}^5,
$$
$$
\bmn=39.70\pm 0.02\times 10^{-4}\,{\rm fm}^5.
$$
In addition, this fit allowed us to determine the values of the effective
radius of the $f_2(1270)$ meson and its decay width into two photons:
$r_f=0.96\pm 0.01$ fm, $\G_{f_2\to\g\g}=3.05\pm 0.11$ keV.

To analyze the results obtained, the DSRs for the difference and sum
of the quadrupole polarizabilities have been constructed.

The values of $\bmn$ and $\bpn$ extracted from the experimental data on
the process $\gg$ are in good agreement with the calculations of the DSRs.

\begin{acknowledgments}
The authors would like to thank D. Hornidge, A.N. Ivanov, M.A. Ivanov,
A.I. L'vov, S. Scherer, and Th. Walcher for useful discussions.
This work was supported in part by the Deutsche Forschungsgemeinschaft
(SFB 443) and the Russian Foundation for Basic Research
(grant No 05-02-04018).
\end{acknowledgments}

\appendix*
\section{}

The contributions of the vector and axial-vector mesons
($\rho$, $\omega$, $\phi$, $a_1$, and $b_1$) are
calculated with the help of the expression
\beq
Im\mpp^{(V)}(s,t)&=&\mp s~Im\mm^{(V)}(s,t)\nonumber \\
&=&\mp 4g_V^2 s\frac{\G_0}
{(m_V^2-s)^2+\G_0^2}
\eeq
where $m_V$ is the meson mass, the sign "+" corresponds to the
 contribution of the $a_1$ and $b_1$ mesons and
$$
g_V^2=6\pi\sqrt{\frac{m_V^2}{s}}\left(\frac{m_V}{m_V^2-\m}\right)^3
\G_{V\to\g\pi},
$$
\be
\G_0=\left(\frac{s-4\m}{m_V^2-4\m}\right)^{\frac32}m_V\G_V.
\ee
Here $\G_V$ and $\G_{V\to \g\pi}$ are the full width and the decay width into
$\g\pi$ of these mesons, respectively.

The contributions of the $a_2$ meson are calculated using
a narrow width approximation.
\beq
Im\mpp^{(a_2)}(s,t)&=&-\frac12 g_a^2 \pi \left[ s^2-s(4\m-2t)+\mu^4\right.
\nonumber \\
&&\left. +\frac{s(s+\m)^2}{2m_a^2}\right]\delta(s-m_a^2) ,
\eeq

\be
Im\mm^{(a_2)}(s,t)=-g_a^2 \pi \left[\m-t-\frac{(s+\m)^2}
{4m_a^2}\right]\delta(s-m_a^2)
\ee
where
$$
g_a^2=160\pi\left(\frac{m_a}{m_a^2-\m}\right)^5\G_{a_2\to \g\pi^{\pm}}.
$$

For calculating the contribution of the $\s$, $f_0(980)$,
$f^{\i}_0(1370)$, and $f_2$ mesons we use the following expressions:
\beq
Im\mpp^{\s}(t,s)&=&\frac{g_{\s}\G_{0\s}}{(m_{\s}^2-t)^2+\G_{0\s}^2},
\nonumber \\
Im\mpp^{f_0}(t,s)&=&\frac{g_{f_0}\G_{0f_0}}{(m_{f_0}^2-t)^2+\G_{0f_0}^2},\\
Im\mm^{f_2}(t,s)&=&\frac{g_{f_2}\G_{0f_2}}{(m_{f_2}^2-t)^2+\G_{0f_2}^2}
\nonumber
\eeq
where
\beq
g_{\s}&=&8\pi\frac{m_{\s}+\sqrt{t}}{\sqrt{t}}\left(\frac{\frac23
\G_{\s\to\pi\pi} \G_{\s\to\g\g}}
{m_{\s}\sqrt{m_{\s}^2-4\m}}\right)^{\frac12}, \nonumber \\
\G_{0\s}&=&\frac{\G_{\s}}2 (\sqrt{t}+m_{\s})\left(\frac{t-4\m}
{m_{\s}^2-4\m}\right)^{\frac12},\\
g_{f_0}&=&16\pi\left(\frac{\frac23 \G_{f_0\to \pi\pi} \G_{f_0\to \g\g}}
{m_{f_0}\sqrt{m_{f_0}^2-4\m}}\right)^{\frac12},  \nonumber\\
\G_{0f_0}&=&\G_{f_0}m_{f_0} \left(\frac{t-4\m}
{m_{f_0}^2-4\m}\right)^{\frac12}, \\
%\eeq
%\beq
g_{f_2}&=&160\pi\frac{m_{f_2}^{3/2}}{t(m_{f_2}^2-4\m)^{\frac54}}
\sqrt{\frac{D_2(m_{f_2}^2)}{D_2(t)}~\G_{f_2\to \pi\pi}
\G_{f_2\to \g\g}}, \nonumber \\
\G_{0f_2}&=&\G_{f_2}\frac{m_{f_2}^2}{\sqrt{t}}\left(\frac{t-4\m}
{m_{f_2}^2-4\m}\right)^\frac52 \frac{D_2(m_{f_2}^2)}{D_2(t)}.
\eeq
The decay form factor $D_2$ is given according to Ref. \cite{mars}
\be
D_2(t)=9+3(q\,r_f)^2+(q\,r_f)^4, \quad q^2=\frac14 (t-4\m)
\ee
where $r_f$ is the effective interaction radius of the $f_2$ meson.
The factor $(m_{\s}+\sqrt{t})$ in the relations for $g_{\s}$ and $\G_{0\s}$
is introduced to get a more correct expression for a broad Breit-Wigner
resonance.


\begin{thebibliography}{99}
\bibitem{ant} Yu.M. Antipov {\em at al.}, Phys. Lett. {\bf B121}, 445 (1983).
\bibitem{pom} I.Ya. Pomeranchuk and I.M. Shmushkevich, Nucl. Phys. {\bf 23},
452 (1961).
\bibitem{star} N.I. Starkov, L.V. Fil'kov, and V.A. Tzarev,
Sov. J. Nucl.Phys. {\bf 36}, 707 (1982).
\bibitem{ayb} T.A. Aybergenov {\em et al.}, Czech. J. Phys. {\bf 36}, 948
(1986).
\bibitem{drech} D. Drechsel and L.V. Fil'kov, Z. Phys. A {\bf 349}, 177 (1994).
\bibitem{ahr} J. Ahrens, V. Alekseyev, J.R.M. Annand {\em et al}.,
Eur. Phys. J. A {\bf 23}, 113 (2005); nucl-ex/0407011.
\bibitem{bab} D. Babusci {\em et al}., Phys. Lett. {\bf B277}, 158 (1992).
\bibitem{don} J.F. Donoghue and B.R. Holstein, Phys. Rev. D {\bf 48}, 137 (1993).
\bibitem{kal} A.E. Kaloshin and V.V. Serebryakov, Z. Phys. C {\bf 32}, 279
(1986); C {\bf 64}, 689 (1994).
\bibitem{ser} A.E. Kaloshin, V.M. Persikov and V.V. Serebryakov,
Phys. At. Nucl. {\bf 57}, 2207 (1994).
\bibitem{mars} H. Marsiske, D. Antreasyan, H.W. Bartels {\em et al}.,
Phys. Rev. D {\bf 41}, 3324 (1990).
\bibitem{fil1} L.V. Fil'kov and V.L. Kashevarov, Eur. Phys. J. A {\bf 5}, 285
(1999).
\bibitem{bij} J. Bijnens and F. Cornet, Nucl. Phys. B {\bf 296}, 557 (1988).
\bibitem{bell} S. Bellucci, J. Gasser, and M.E. Sainio, Nucl. Phys.
B {\bf 423}, 80 (1994).
\bibitem{oller} J.A. Oller and E. Oset, Nucl. Phys. A {\bf 629}, 739 (1998).
\bibitem{lee} C.-H. Lee, H. Yamagishi, and I. Zahed, Nucl. Phys. A {\bf 653},
185 (1999).
\bibitem{rad1} I. Guiasu and E.E. Radescu, Ann. Phys. (N.Y.) {\bf120}, 145
(1979).
\bibitem{lvov} D. Babusci, G. Giordano, A.I. L'vov, G. Matone, and
A.M. Nathan, Phys. Rev. C {\bf 58}, 1013 (1998).
\bibitem{holst} B.R. Holstein, D. Drechsel, B. Pasquini, and
M. Vanderhaeghen, Phys. Rev. C {\bf 61}, 034316 (2000).
\bibitem{fil2} L.V. Fil'kov, Sov. J. Nucl. Phys. {\bf 41}, 636 (1985).
\bibitem{bien} J.K. Bienlein, Crystal Ball Contribution to the 9th Intern.
Workshop on Photon-Photon Collisions, La  Jolla, 23--26 March 1992;
ISSN 0418-9833.
\bibitem{aber} H.A. Abarbanel and M.L. Goldberger, Phys. Rev. {\bf 165},
1594 (1968).
\bibitem{pran} R. Prange, Phys. Rev. {\bf 110}, 240 (1958).
\bibitem{guias} I. Guiasu and E.E. Radescu, Ann. Phys. (N.Y.) {\bf 122}, 436
(1979).
\bibitem{rad} L.V. Fil'kov, I. Guiasu, and E.E. Radescu, Phys. Rev. {\bf 26},
3146 (1982).
\bibitem{pdg} S. Eidelman {\em et al.}, (PDG), Phys. Lett. B {\bf 592}, 1
(2004).
\bibitem{anis} V.V. Anisovich {\em et al.}, Phys. At. Nucl. {\bf 65},
1545 (2002).
\bibitem{ait} E.M. Aitala {\em et al.}, Phys. Rev. Lett. {\bf 86}, 765 (2001).
\bibitem{morg} D. Morgan and M.R. Pennington, Z. Phys. {\bf C48}, 623 (1990).
\bibitem{bugg} D.V. Bugg, A.V. Sarantsev, and B.S. Zou, Nucl. Phys.
{\bf B471}, 59 (1990).
\bibitem{zel} M. Zielinski {\em et al.}, Phys. Rev. Lett. {\bf 52}, 1195
(1984).
%\bibitem{proc} T.A. Aybergenov {\em et al.}, Proceedings of the Lebedev
%Phis. Inst. {\bf 186}, 169 (1998).
%\bibitem{kam} R. Kaminski, L. Lesniak, and B. Loiseau ,
%Eur. Phys. J C {\bf 9}, 141 (1999).
\bibitem{burgi} U. B\"urgi, Nucl. Phys. B {\bf 479}, 392 (1997).
\bibitem{gray} G. Grayer {\em et al.}, Nucl. Phys. B {\bf 75}, 189 (1974).
\end{thebibliography}
\end{document}